\documentclass[sigconf]{acmart}

\AtBeginDocument{%
  \providecommand\BibTeX{{%
    \normalfont B\kern-0.5em{\scshape i\kern-0.25em b}\kern-0.8em\TeX}}}

\setcopyright{acmcopyright}
\copyrightyear{2018}
\acmYear{2018}
\acmDOI{10.1145/1122445.1122456}

\acmConference[Woodstock '18]{Woodstock '18: ACM Symposium on Neural
  Gaze Detection}{June 03--05, 2018}{Woodstock, NY}
\acmBooktitle{Woodstock '18: ACM Symposium on Neural Gaze Detection,
  June 03--05, 2018, Woodstock, NY}
\acmPrice{15.00}
\acmISBN{978-1-4503-XXXX-X/18/06}

\copyrightyear{2020} 
\acmYear{2020} 
\setcopyright{acmcopyright}\acmConference[ICAIF '20]{ACM International Conference on AI in Finance}{October 15--16, 2020}{New York, NY, USA}
\acmBooktitle{ACM International Conference on AI in Finance (ICAIF '20), October 15--16, 2020, New York, NY, USA}
\acmPrice{15.00}
\acmDOI{10.1145/3383455.3422535}
\acmISBN{978-1-4503-7584-9/20/10}

\begin{document}

\title{What can be learned from satisfaction assessments?}

\author{Naftali Cohen}
\authornotemark[1]
\affiliation{%
  \institution{J.~P.~Morgan AI Research}
  \city{New York}
  \state{NY}}
\email{naftali.cohen@jpmchase.com}

\author{Simran Lamba}
\affiliation{%
  \institution{J.~P.~Morgan AI Research}
  \city{New York}
  \state{NY}}
\email{simran.lamba@jpmchase.com}

\author{Prashant Reddy}
\affiliation{%
  \institution{J.~P.~Morgan AI Research}
  \city{New York}
  \state{NY}}
\email{prashant.reddy@jpmchase.com}

\renewcommand{\shortauthors}{Cohen N., et al. 2020}

\begin{abstract}
Companies survey their customers to measure their satisfaction levels with the company and its services. The received responses are crucial as they allow companies to assess their respective performances and find ways to make needed improvements. 
This study focuses on the non-systematic bias that arises when customers assign numerical values in ordinal surveys.   
Using real customer satisfaction survey data of a large retail bank, we show that the common practice of segmenting ordinal survey responses into uneven segments limit the value that can be extracted from the data.
We then show that it is possible to assess the magnitude of the irreducible error under simple assumptions, even in real surveys, and place the achievable modeling goal in perspective.   
We finish the study by suggesting that a thoughtful survey design, which uses either a careful binning strategy or proper calibration, can reduce the compounding non-systematic error even in elaborated ordinal surveys.
A possible application of the calibration method we propose is efficiently conducting targeted surveys using active learning.
\end{abstract}



\maketitle

\section{Introduction}
The Net Promoter Score (NPS) is an index ranging between -100 to 100 and is used as a proxy for assessing overall customer satisfaction and loyalty to a company or its services. It is considered to be the single most reliable indicator of a firm's growth compared to other loyalty metrics, such as customer satisfaction [e.g., \citeauthor{reichheld2003one}, \citeyear{reichheld2003one}; \citeauthor{farris2010marketing}, \citeyear{farris2010marketing}].
NPS is widely adopted by thousands of  well-established companies, including Amazon, Apple, Netflix, Walmart, and Vanguard
[e.g., \citeauthor{reichheld2011ultimate}, \citeyear{reichheld2011ultimate}; \citeauthor{satmetrix2018}, \citeyear{satmetrix2018}].
To calculate the NPS, customers are asked to answer a single question similar to the following: "\emph{On a scale of 1-10, how likely are you to recommend our brand to a friend or colleague?}"
Customers who respond with a score of 9 or 10 are classified as Promoters, responses of 7 and 8 are labeled as Passives/Neutrals, and those who give a score of 1 to 6 are called Detractors. The NPS is then calculated by subtracting the percentage of detractors from the percentage of promoters.

NPS vary widely by industry 
[e.g., \citeauthor{reichheld2011ultimate}, \citeyear{reichheld2011ultimate}]. For example, in a 2018 study published by NICE Satmetrix [\citeauthor{satmetrix2018}, \citeyear{satmetrix2018}], the average NPS of the airlines industry was 44, while for the health insurance sector, it was only 13. However, per sector, if a company has a substantially higher NPS than its competitors, it is likely to grow at a much faster rate than its rivals [e.g., \citeauthor{reichheld2003one}, \citeyear{reichheld2003one}].

Each company's actual NPS is unknown, but an approximation can be computed via surveys [e.g., \citeauthor{groves2011survey}, \citeyear{groves2011survey}]. In reality, however, survey results must be considered with care due to a variety of systematic and non-systematic biases as coverage error, sampling error, nonresponse error, measurement error, and random error [e.g., \citeauthor{furnham1986response}, \citeyear{furnham1986response}; \citeauthor{weisberg2009total}, \citeyear{weisberg2009total}]. 

This paper assumes that all systematic bias is negligible and focuses our attention on non-systematic bias.
For this study, we define each respondent's
opinion in terms of a probability distribution. 
That is, if respondents always respond to the same questions in the same way, it means that their opinion distributions follow a delta function that is centered at their true opinion. Practically, it is more natural to relax this assumption and assume that on average respondents have consistent opinions, but their opinions follow wider probability distributions. 
For example, consider a genuine promoter of a brand. In the framework of the above question, this customer will respond with a 10 in the ordinal survey.
However, if we give this customer an infinite number of surveys with the same exact question, will they mark 10 each time? In this study, we think of customers as individuals who, on average, express their opinions consistently. We ask what is the effect of sampling, in a given survey, from the population of each respondent's opinion distribution? Sampling from each person's opinion invokes the notion of inherent variability. We want to examine whether this noise cancels out or compounds, in particular when we use the common practice of unevenly-spaced binning of ordinal responses.
In the following examples, the survey responses are used to label the data, and the focus is on the effect of learning from non-systematic noisy labels.

\section{Related Work and Main Contribution}
The problem of learning from noisy labels has been widely addressed in recent years [e.g., \citeauthor{frenay2013classification}, \citeyear{frenay2013classification}; \citeauthor{natarajan2013learning}, \citeyear{natarajan2013learning}; \citeauthor{weisberg2009total}, \citeyear{weisberg2009total}], where in most cases, the causes for the ill-labeling include: insufficient information for proper labeling, human error, and even malicious attacks. Most frequently, the  recognized consequence of noisy labels is a decrease in classification performance and data reliability [e.g., \citeauthor{quinlan1986induction}, \citeyear{quinlan1986induction}; \citeauthor{weisberg2009total}, \citeyear{weisberg2009total}; \citeauthor{nettleton2010study}, \citeyear{nettleton2010study}], changes in learning requirements [e.g., \citeauthor{angluin1988learning}, \citeyear{angluin1988learning}], increase in the complexity of learned models [e.g., \citeauthor{zhu2004class}, \citeyear{zhu2004class}], distortion of observed frequencies [e.g., \citeauthor{van2002randomized}, \citeyear{van2002randomized}], difficulties in identifying relevant features [e.g., \citeauthor{shanab2012robustness}, \citeyear{shanab2012robustness}], and many more. 
In this study, we focus on the decrease in classification performance and distortion of observed frequencies solely due to random noise in the labels. 
The central approaches [e.g., \citeauthor{frenay2013classification}, \citeyear{frenay2013classification}] to deal with the problem use algorithms that are more robust to label noise and involve identifying and filtering the noisy labels out before training [e.g., \citeauthor{garcia2019enabling}, \citeyear{garcia2019enabling}], or even directly modeling the noise in labels during learning.

This work concentrates on data and on the standard industry practice of measuring and assessing customer satisfaction using NPS index. We focus attention on the case where the NPS survey responses are ordinal and segmented into unevenly-spaced bins. 
The main contribution of this study is that we demonstrate how the (almost exponential) decrease in the classification performance can be estimated in real data. We also show that the various bin designs can have a cost up to 20\% in accuracy scores. Lastly, we propose a simple solution to reduce the non-systematic noise in survey response data by adding a short textual description to the numerical ratings.

\begin{figure}[t]
\centering
\noindent\includegraphics[width=0.75\columnwidth]{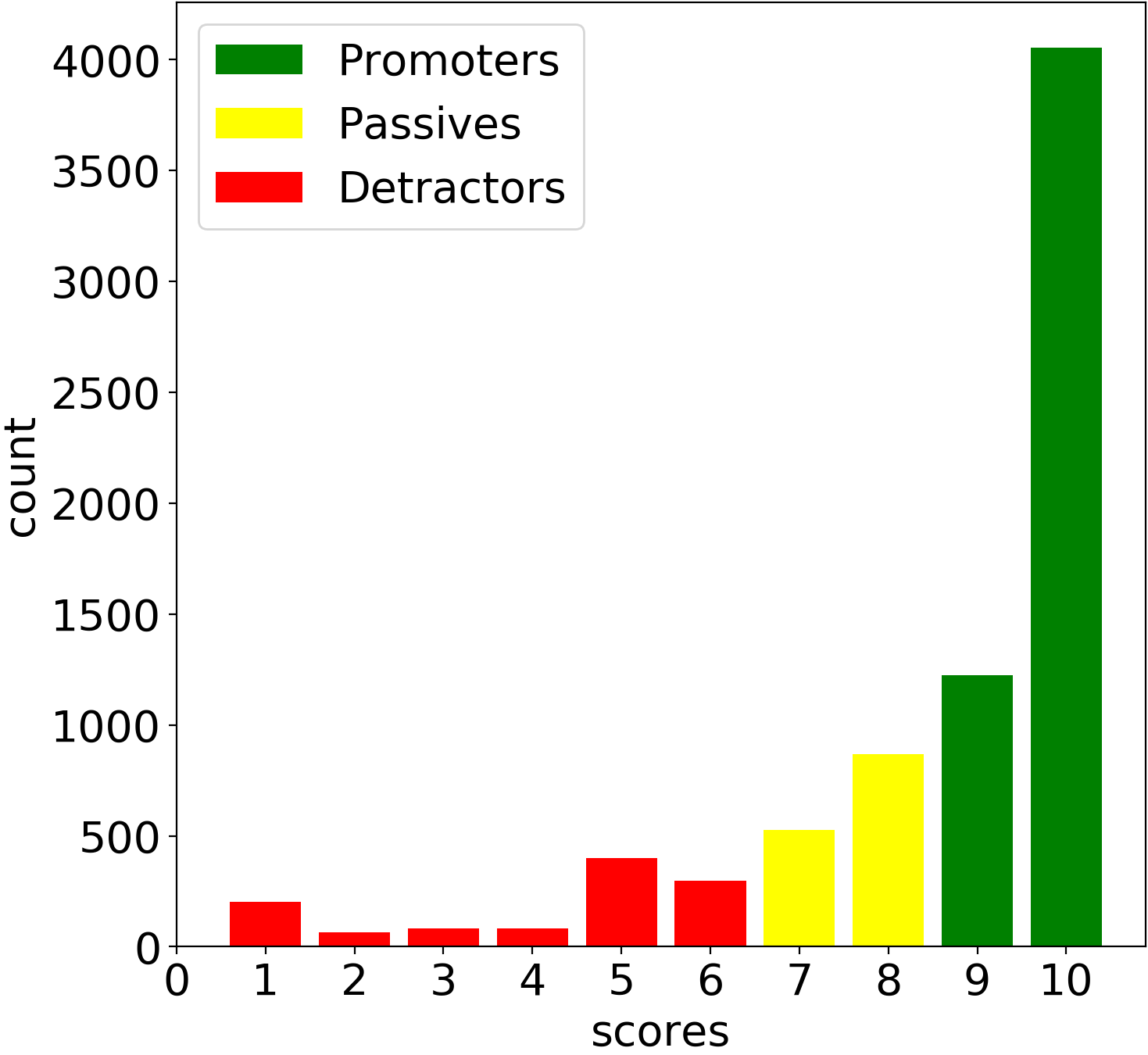}
\caption{The brand satisfaction survey data are highly imbalanced.}
\end{figure}

\section{Data and Methods}
In this section, we explore the three data sets that are used in this study and the methods we take to explore them.

\begin{enumerate}
\item The first is a real NPS satisfaction survey data of a large retail bank collected during December 2018 (hereafter, BRAND). The survey aims at measuring the overall satisfaction of the customers toward BRAND. For that measure, customers were asked: "Would you recommend BRAND to a friend or colleague? Please use a scale of 1 to 10, where 1 is \emph{Definitely Not}, and 10 is \emph{Definitely}."

The BRAND data we use in this study includes the response of 10,000 unique customers. Besides the numeric responses to the survey, each customer is characterized by numerous demographics and product usage features.

Figure 1 shows the distribution of the survey scores. Clearly, the data is left-skewed and highly imbalanced by score. It is almost log-normally distributed: most of the customers surveyed
gave a score of 10, following by customers who gave scores of 9 and 8. Only a few gave scores of 1 and 5 to 7, while even less chose to give scores of 2 to 4.

\begin{figure}[t] 
\centering
\noindent\includegraphics[width=1.0\columnwidth]{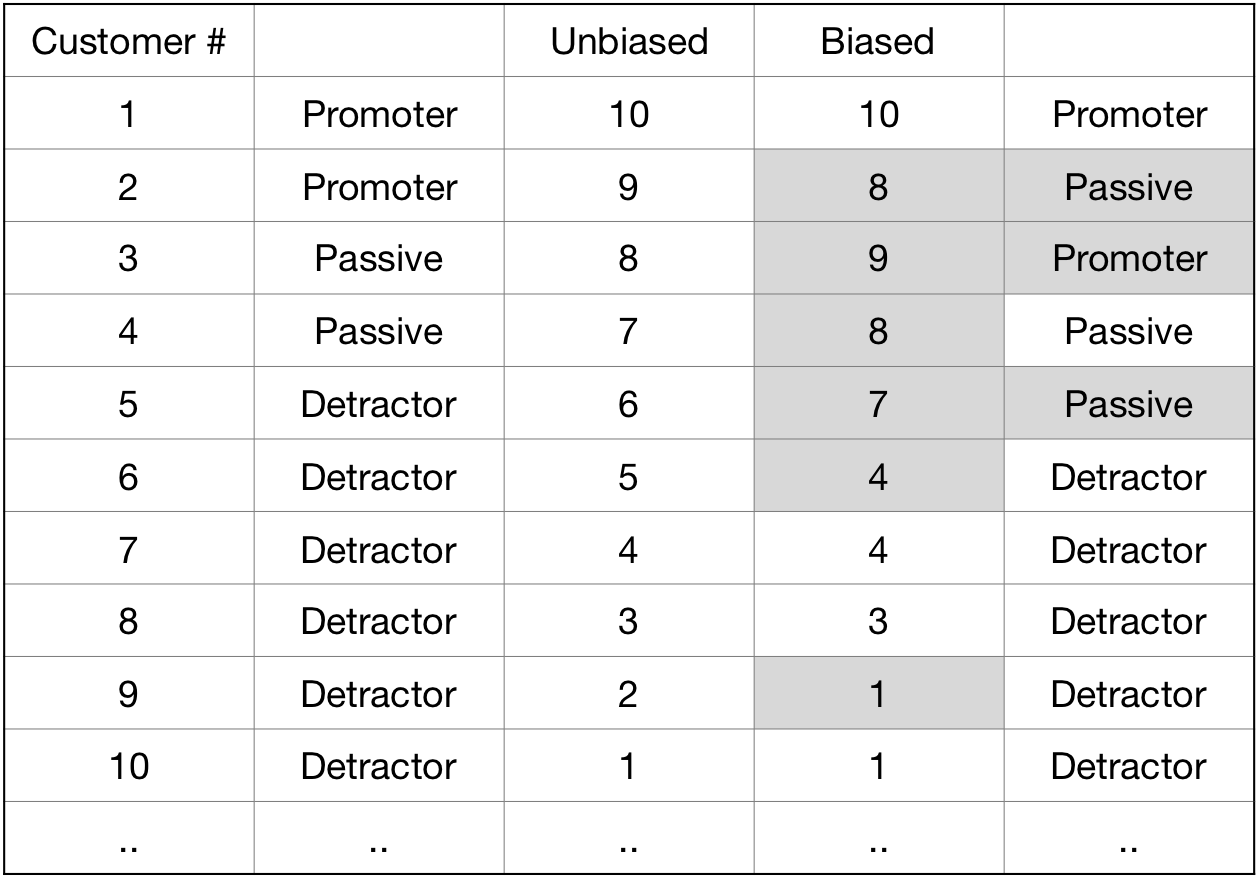}
\caption{Unbiased and biased survey responses.}
\end{figure}

Keep in mind that, following the BRAND's business model, each survey score gets categorized in the following way: customers who gave a score of 9-10 are considered "Promoters," those who gave a score of 7-8 are considered "Passives," while customers who picked 1-6 are considered "Detractors." Figure 1 distinguishes between the various categories following the color code, as shown in the figure's legend.

The percentages of customers in the Promoter and Detractor categories are then used to compute the overall Net Promoter Score (NPS) of the brand in the following way: 
\begin{equation}
\textrm{NPS}=\%\textrm{Promoters}-\%\textrm{Detractors}.    
\end{equation}

The NPS metric varies between 100 to -100 and is considered critical as it is argued to be positively correlated with the future success of the brand, as discussed in the Introduction section.
Using the data presented in Fig.~1, BRAND gets the very high NPS of 53.

\item The second data set is a synthetically generated data (hereafter, SYNTH). To create this data, we consider 10,000 customers, each given a score that varies between 1 to 10, and that is drawn randomly from a uniform distribution. The Unbiased column in Fig.~2 shows ten such customers and their corresponding categories following the above BRAND definition.

A key assumption in this study is that each customer has an \emph{unbiased} (or systematic averaged) opinion. In an example, the true satisfactory level of the customer at the first row has the value of 10, and the customer at the second row has a true satisfactory level of 9. Both of them are genuine promoters (by category) of the BRAND. 

A second important assumption in this study is that, in a survey, people might express a different opinion than their unbiased, true one.
In other words, there is an intrinsic non-systematic bias in the way people express themselves in surveys. If the intrinsic variability is zero, the score a customer specifies in a survey always equals their true opinion. For example, if the above two customers with the unbiased satisfaction levels of 10 and 9 have zero intrinsic variability, then their survey responses will always be 10 and 9, respectively.

\begin{figure}[t] 
\centering
\noindent\includegraphics[width=0.754\columnwidth]{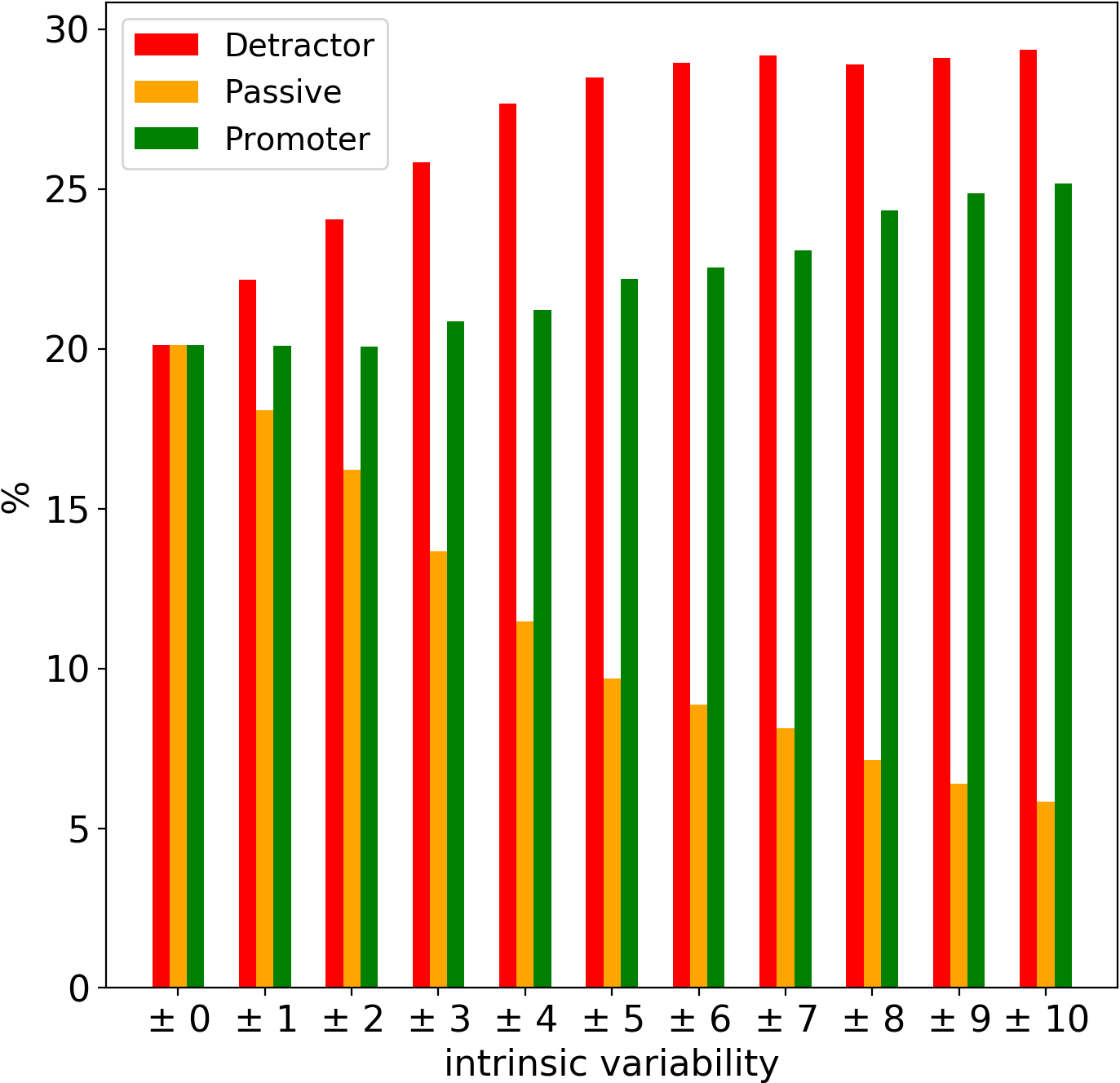}
\caption{The effect of intrinsic variability on class imbalance.}
\end{figure}

In this paper, for simplicity, we assume that the intrinsic variability follows a discrete uniform distribution.
The column Biased in Fig.~2 shows the case of a uniformly-distributed intrinsic variability of $\pm1$ about their Unbiased score. In that case, a customer with a true satisfaction level of 9 can equally likely mark an 8, 9, or 10 in a survey, whereas a person with a true opinion of 6 may, similarly, give a score of 5, 6, or 7. Because of the upper bound of 10 in the survey score, a person with an unbiased satisfaction level of 10 and an intrinsic variability of $\pm1$ might give a 9, 10, or 10 (and similarly for when encountering the lower bound): this individual has a 2/3 chance of stating 10, while only a 1/3 chance of stating 9 in a survey. Mathematically, this is formulated by applying a simple min-max operator on the scores that are drawn from the discrete uniform distribution.

The gray shading in the column Biased in Fig.~2 represents customers whose scores changed because of the above-mentioned reason. The right-most column shows the categories that correspond to the Biased scores. It can be seen that some (but not all) of the Biased scores that are marked in gray do not match with their original category.

\item The third data set comes from an online survey we conducted particularly for this study (hereafter, CITY). We surveyed about 200 employees of the BRAND and examined the degree of non-systematic error in their responses to ordinal surveys. The design of the survey is such that along with the biased responses, we also collect an approximation to the true underlying unbiased opinions. 

The survey starts with the following question, "In what city do you live?" In the following questions, the participants are asked to enumerate and categorize their satisfaction level in reference to the city they stated.

\begin{figure}[t]
\centering
\noindent\includegraphics[width=0.754\columnwidth]{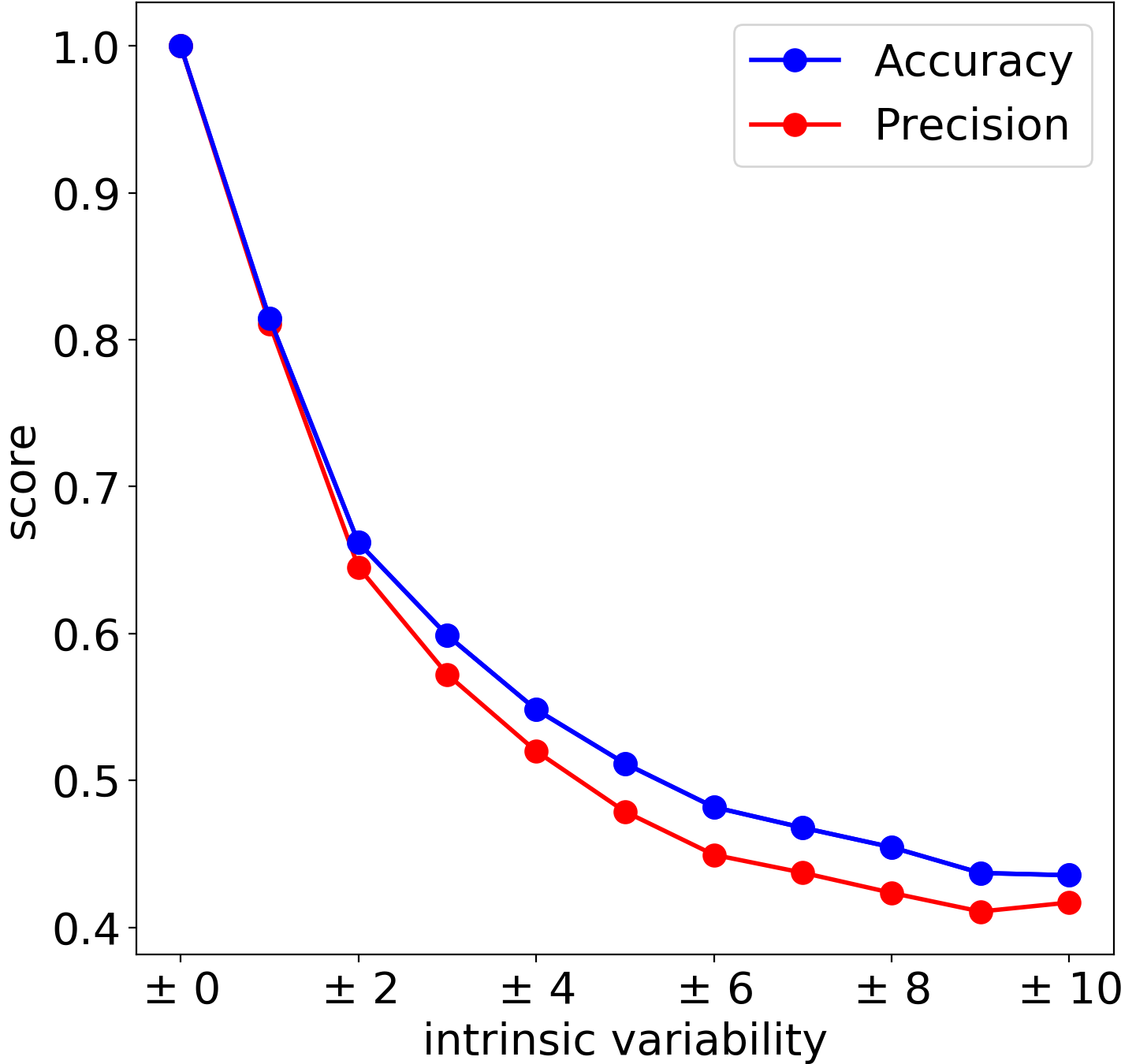}
\caption{Upper bound for three-class classification as a function of the intrinsic variability.}
\end{figure}

In the second question, we ask the participants to assign an ordinal score to their general satisfaction toward the city they chose by asking, "Rate your city as a place to live on a scale of 1-10." The next question asks them to self-assign a matching category by asking, "My city is a \_\_\_\_ city to live in," where they had to choose among the three possible categories "great," "okay," or "bad."
This question aims at quantifying whether the BRAND'S categories that are binned 1-6, 7-8, and 9-10, are natural to survey participants. In other words, it seeks to identify whether people who replied 1-6 on question 2 would mark "bad," whether those who responded 7-8 would mark "okay," and whether those who answered 9-10 would mark "great."

\begin{figure}[t]
\centering
\noindent\includegraphics[width=0.754\columnwidth]{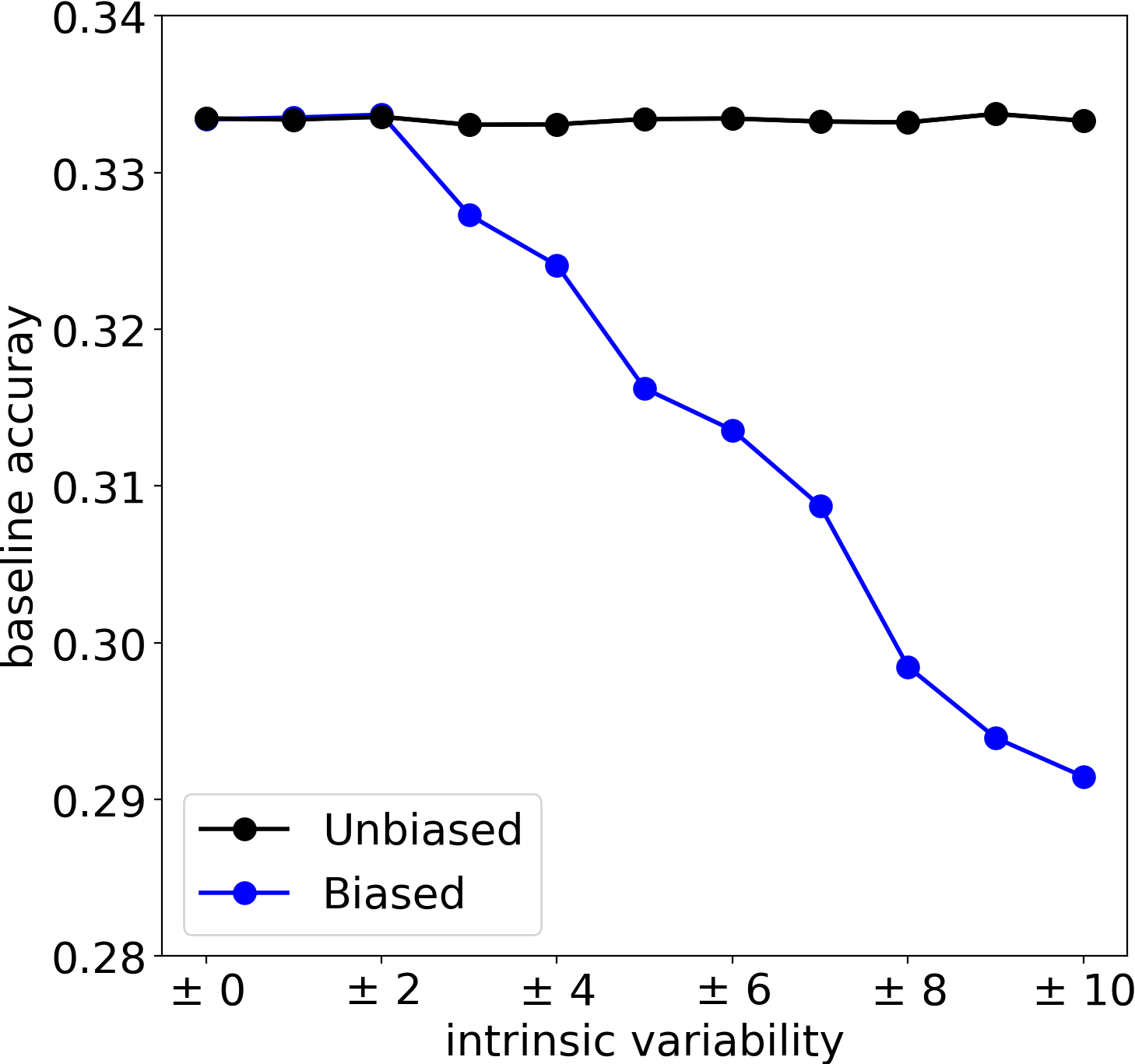}
\caption{Lower bound for three-class classification as a function of the intrinsic variability.}
\end{figure}

The primary problem with ordinal surveys that measure subjective opinions (i.e., satisfaction level) is that they are not calibrated. In the next two questions, we try to have a sense of the upper and lower bounds of the true underlying scales of the survey respondents. 
To achieve that we ask explicitly "What is the highest rate you would ever give in a survey like this?" and "What is the lowest rate you would ever give in a survey like this?"

We conclude the study CITY with a proposition for a semi-ordinal, text-calibrated survey that is used as an approximation to the true underlying unbiased opinions. We repeat the second survey question asking to rate the city on an ordinal scale, but here each numeric rating on the scale has a short description attached to it. 
In an example, the choice "7" is replaced by "7) My city is a decent place," and "8" by "8) My city is very nice," etc. 
This allows the participants to calibrate their responses not only to ordinal scales but also to a universal textual description of what each numerical category means.

\end{enumerate}

\section{Results}
In this section, we report the results from the analysis of the various data sets mentioned above. We start by examining the SYNTH data. To create this data, as mentioned in the previous section, we sample at random 10,000 scores from a discrete uniform distribution that spans from 1 to 10. In terms of BRAND's categorization, the classes are imbalanced because scores between 9-10 are assigned to the Promoters group, while 7-8 to the Passives, and 1-6 to Detractors. 

Figure 2 shows an example of ten customers with a uniformly distributed intrinsic variability of $\pm1$ about their unbiased score. The Unbiased column refers to the customers' true (or systematically averaged) opinion, while the Biased column shows a sampled example of what they might have marked in a particular survey. 

The accuracy that corresponds to the Biased column relative to the Unbiased column is 0.4. However, the accuracy of the corresponding categories is 0.7, which is much higher due to the binning effect. 
The problem with the data in Fig.~2 is that the survey scores and corresponding categories are not balanced, which makes the structural learning of each class uneven. Also, this makes the interpretation of the accuracy score nonintuitive. 
To workaround that challenge we follow the undersampling method [e.g., \citeauthor{he2008learning}, \citeyear{he2008learning}; \citeauthor{fernandez2018learning}, \citeyear{fernandez2018learning}] and balance the classes by repeatedly sampling\footnote{we repeatedly sample both classes 1000 times, where repetitions are allowed, according to the size of the minority class. Then we report on the ensemble-mean metrics.} the minority and majority classes according to the size of the minority class.
In the example of Fig.~2, this results in considering only two customers per category each time. The average category accuracy over the balanced set is 0.6 -- a slight decrease that reflects the fact that there are more mismatches in the minority categories. A side effect with far-reaching implications of the inherent variability is that the biased categories go more and more unbalance for increased variability as we are about to show. 

\begin{figure}[t] 
\centering
\noindent\includegraphics[width=0.754\columnwidth]{classification_analysis_2.png}
\caption{The effect of inherent variability on the NPS scores.}
\end{figure}

Figure 3 shows how the class imbalance of the \emph{biased} category in the SYNTH data develops as a function of the intrinsic variability. 
When the intrinsic variability equals zero, all three classes are equal in size. In other words, each class takes about 20\% of the whole data as can be seen on the y-axis of Fig.~3. As the intrinsic variability increases, the Detractor category grows in size at a faster rate than the Promoter category. At the same time, the class of Passives shrinks dramatically. This effect is seen because the increased uniform variability accumulates the scores more at the high and low categories than the middle one.

A key point in this study is to evaluate the effect of the intrinsic variability on the upper-bound classification scores. 
Figure 4 shows how the intrinsic variability affects the category classification accuracy and precision scores as the variability increases. 
When the intrinsic variability is  zero, the data is balanced and the upper bounds for both accuracy and precision scores stand at 1. However, as the variability increases the upper bounds decrease dramatically. It is important to note that even for the smallest variability of $\pm1$ there is a 20\% decrease in the accuracy and precision upper bounds -- from 1 to about 0.8.

Figure 5 shows, in a similar manner, how the lower bound of the three-class classification accuracy change as the intrinsic variability increases. The Unbiased score (in black) stands at 1/3, which is equal to a random guess over a balanced set of three classes.
On the other hand, the lower bound on the Biased scores (in blue) do change as the intrinsic variability increases. This unexpected effect can be traced back to the drastic shrinkage in the Passive category relative to the other categories, as seen in Fig.~3.

\begin{figure}[t]
\centering
\noindent\includegraphics[width=0.754\columnwidth]{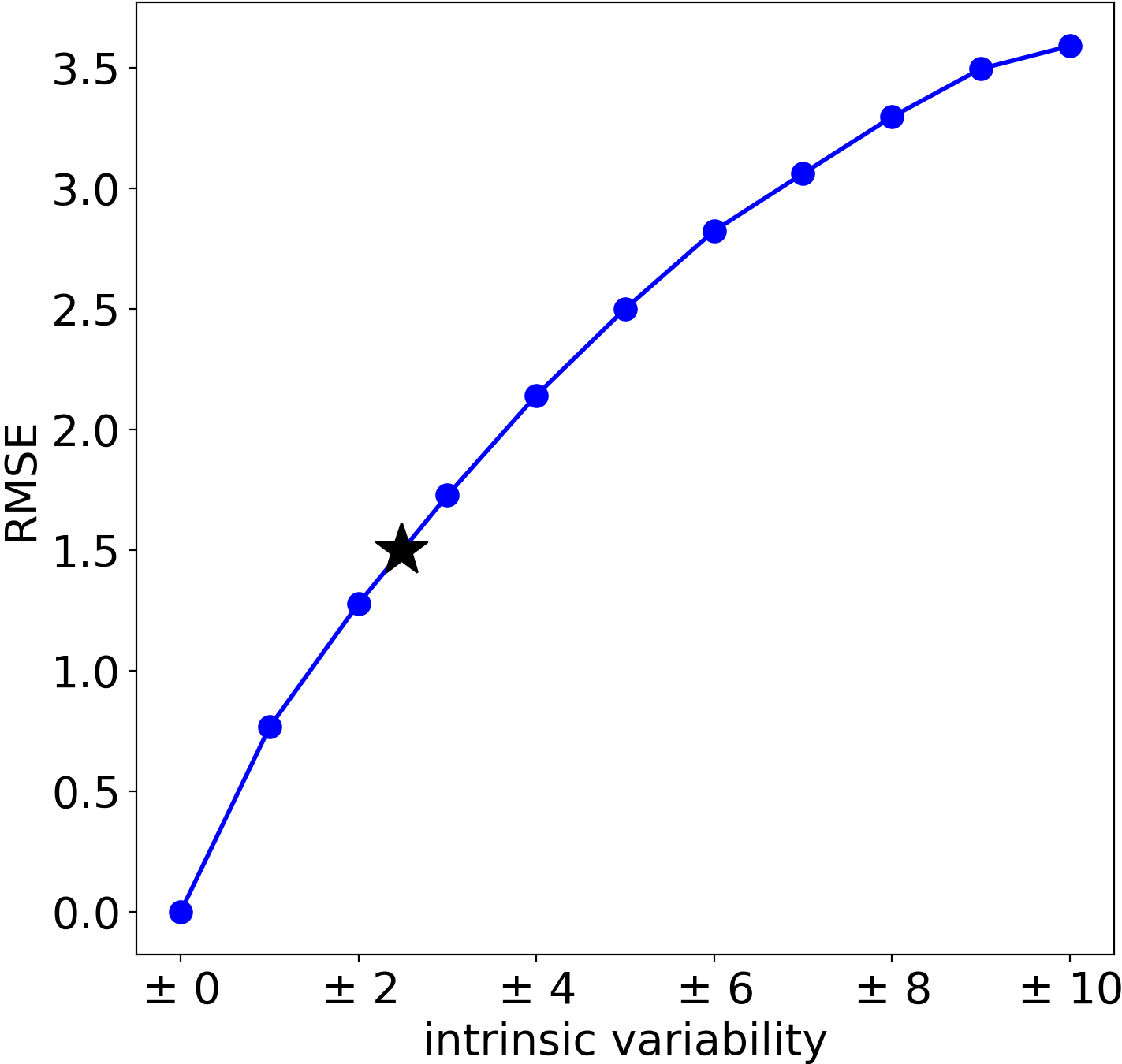}
\caption{Regressing Biased on Unbiased survey scores.}
\end{figure}

The curves in Figs.~4 and 5 mark the upper and lower bounds for the achievable accuracy and precision scores for the various intrinsic variabilities. In other words, the consequence of uneven binning over noisy ordinal labels is that there is a substantial limitation on the predictability that one can expect to extract from the binned data. The key here is that for binned ordinal labels, the non-systematic error does not cancel out but, instead, accumulates and compounds.

Figure 6 shows how the binned inherent variability affects the NPS scores. When the intrinsic variability is zero, there is no change to the initial NPS score. As the variability increases, the NPS score starts decreasing, reaching a maximum change of about -10\% at intrinsic variability of $\pm4$. However, compared to the change in accuracy scores, there is only a minimal change in the initial NPS, and also the decrease is not monotonic. The reason for this is seen in Fig.~3 -- as the variability increases, there is a pronounced gap between the Detractors and Promoters classes. However, this gap maintains almost a constant value. In comparison, the group of Passives decreases dramatically, relative to the Detractors and Promoters classes. 
In other words, the class imbalance affects the gap accumulation between Passives and the other two categories -- decreasing the accuracy and precision scores. On the other hand, the NPS accounts for the difference between the Promoter and Detractor classes, and these classes maintain relatively stable ratio even as the intrinsic variability increases.

Figure 7 shows one of the novel results of this study by showing how to estimate the intrinsic variability in real ordinal survey data, (i.e., the BRAND data) under the uniform distribution assumption. 
Looking back at the ten customers example in Fig.~2, one can work out not only the classification problem but also the regression problem: consider the Unbiased score column as the independent variable (say, $x$) and the Biased score column and the dependent variable (say, $y$). Then, solve the linear-regression problem [e.g., \citeauthor{wilks2011statistical}, \citeyear{wilks2011statistical}] by simply regressing $y$ on $x$, or the Biased scores on the Unbiased scores. 

\begin{figure}[t] 
\centering
\noindent\includegraphics[width=0.754\columnwidth]{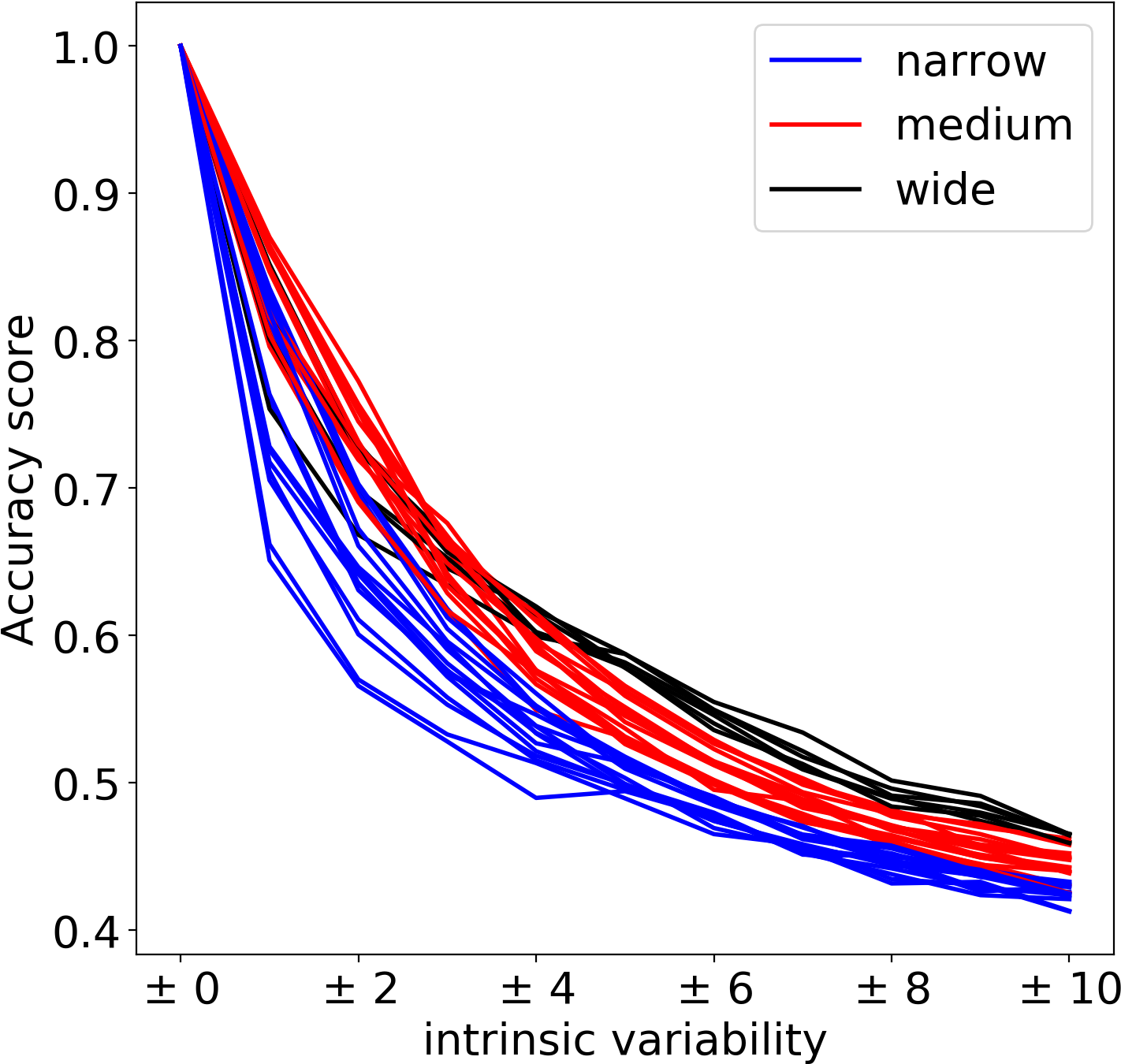}
\caption{The effect of three-class category design on the achievable accuracy.}
\end{figure}

Figure 7 shows the results of such an experiment applied to the SYNTH data as a function of the increased variability. As expected, the root-mean-square error (hereafter, RMSE), increases as the inherent variability increases. 
Similarly, one can work out the multivariate linear regression problem and best fit a multi-feature real ordinal survey data (i.e., the BRAND data) to its labels (i.e., the survey scores). The key idea is that one can then use the RMSE of the real data (after balancing it), equate it to the RMSE of the synthetic data, and read off the inherent variability of the real data from Figure 7. This is important as the inherent variability puts an upper bound on the achievable accuracy and predictability skill in the data, as shown in Figure 4.
For the BRAND data, this procedure results in estimating the inherent variability to be around $\pm2.5$ (marked by a black star), which means that the upper bound on accuracy scores for the three-class category classification data stands at about 0.65 (see Fig.~4).

\begin{figure}[t] 
\centering
\noindent\includegraphics[width=0.754\columnwidth]{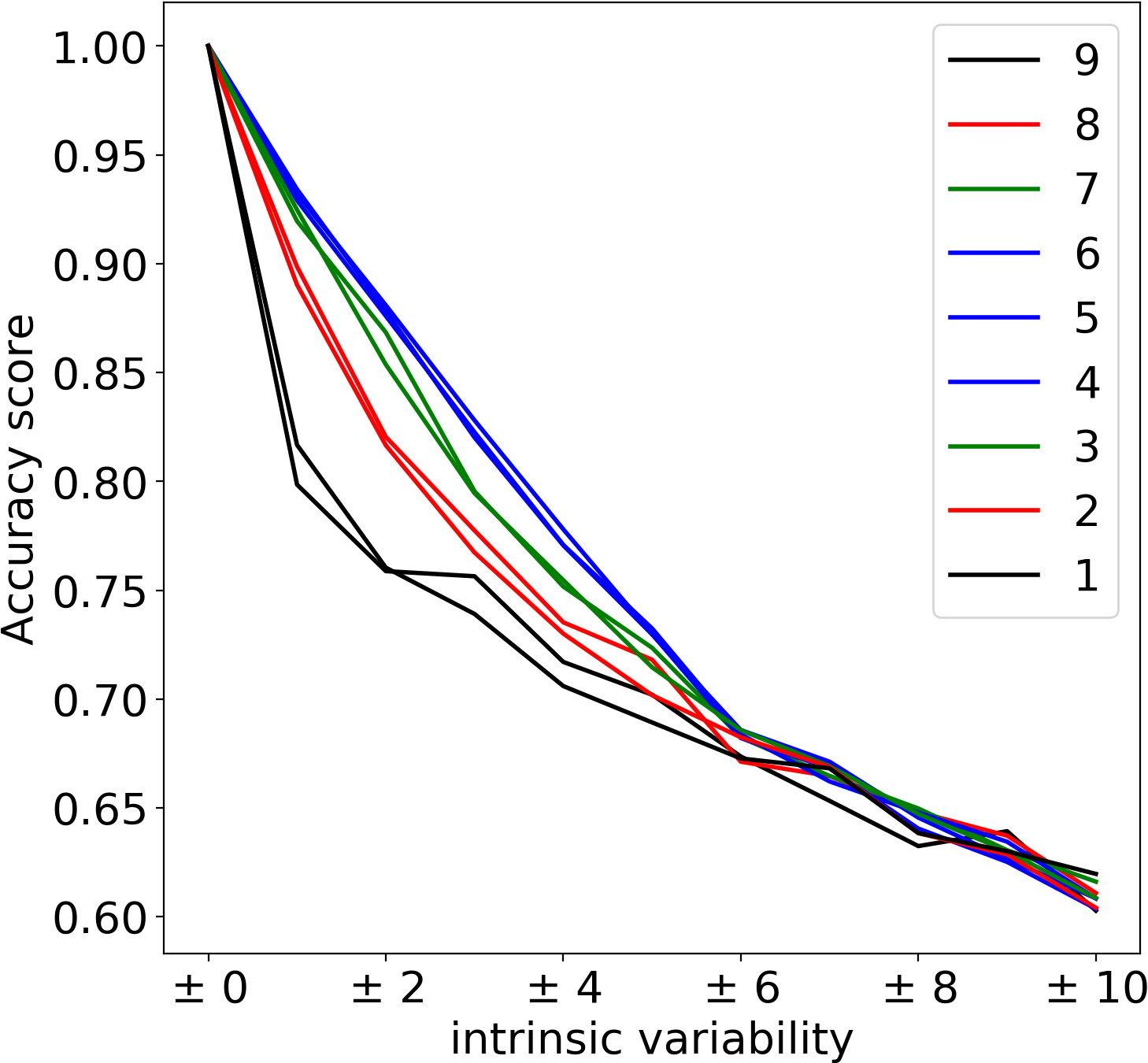}
\caption{The effect of two-class category design on the achievable accuracy.}
\end{figure}

\section{Discussion}
We finished the previous section by arguing that the uniform intrinsic variability assumption allows one to relate and estimate the variability in real surveys and that this puts an estimated upper bound on the achievable real data classification metrics. 

The practical and overlooked consequence is that there is a difference between the actual classification score that one can extract from the data using machine-learning classification algorithms and the effective score relative to its upper bound. For example, if one works out the balanced three-class classification problem of the BRAND data and achieves an actual accuracy of 0.55, when put in perspective of its upper bound it means that the relative accuracy of the data is 0.55/0.65 or 0.85 -- a number that is almost twice as large as the raw accuracy. 
In other words, the accuracy is still 0.55, but relative to the amplitude of noise in the ordinal labels, the model is able to extract most, or 85\%, of the predictability in the data.

Up until now, we discussed the case where the categories were decided using the BRAND's unevenly-spaced binning. However, is there a better binning design for the ordinal scores to minimize the effect on the non-systematic error accumulation? 
To answer that question, we conduct an experiment where we consider all the ways by which one can split the ten ordinal scores into three categories. In an example, one split can be [1-3, 4-6, 7-10], while another can be [1-3, 4-7, 8-10], or [1-6, 7-8, 9-10] as in BRAND's business rule. An easy calculation shows that there are 45 ways to split the scores into three bins. 
To summarize the performances for each possible split, we compute the length of the middle class. For example, the three designs mentioned above will get the values of 3, 4, and 2, respectively. 

\begin{figure}[t] 
\centering
\noindent\includegraphics[page=2,width=0.754\columnwidth]{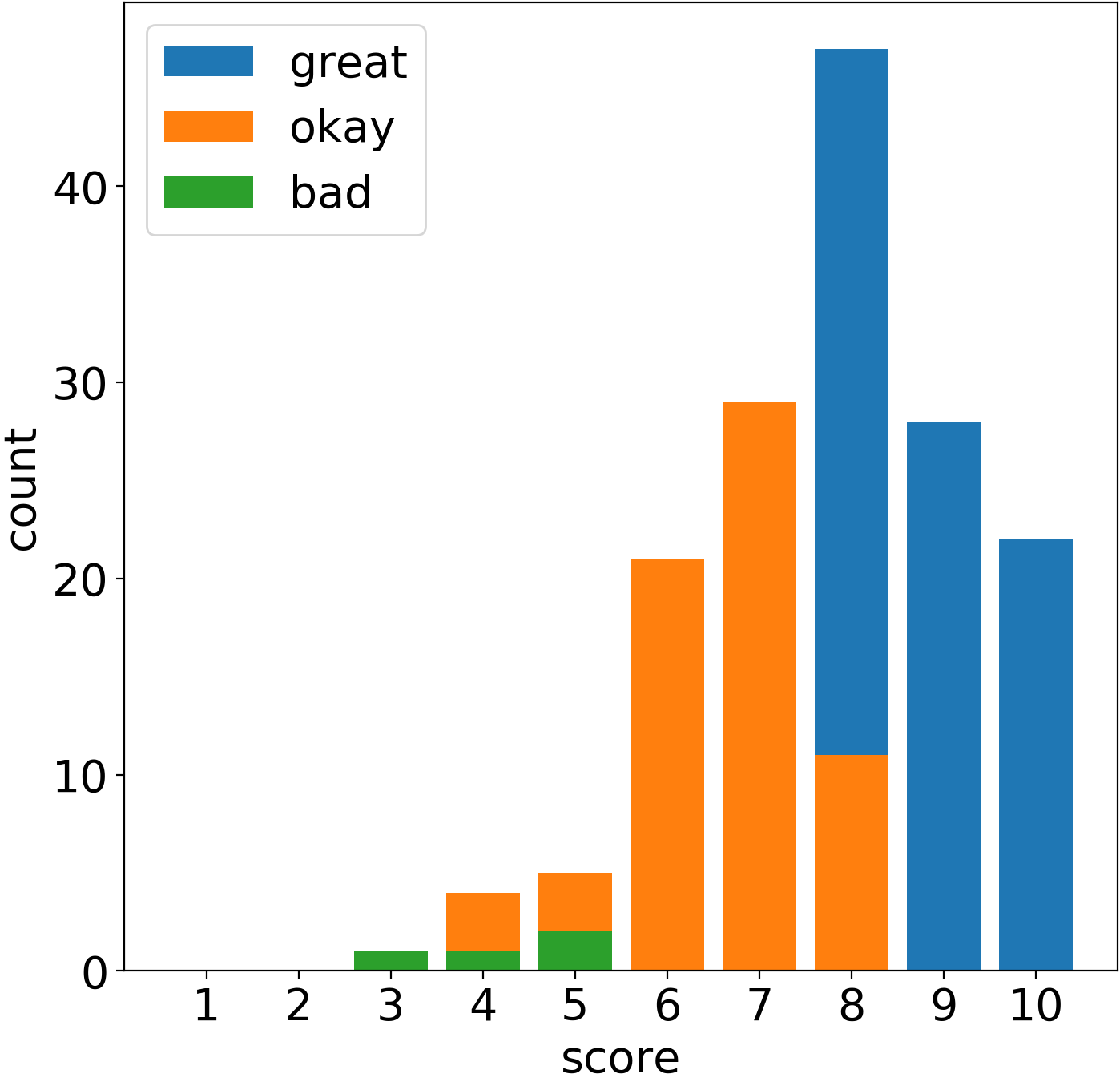}
\caption{Survey responses and the natural assignment to categories.}
\end{figure}

Figure 8 shows the results of the analysis where "narrow" denotes middle-class lengths of less than 3, "medium" denotes lengths between 3 and 5, and "wide" denotes lengths of 6 and above. The figure shows that for variability values at or less than $\pm3$, the best configuration is "medium" while for variability above $\pm3$, the best configuration is "wide." For all cases, the worst configuration is "narrow."
The intuition behind this result is quite simple: for "narrow" middle-class configuration, even small variability causes significant leakage from the middle category, which reduces accuracy scores. On the other hand, for high variability, the best configuration is "wide" because the broad middle category remains relatively untouched while the upper and lower classes accumulate as well. For low variability, the "medium" configuration is best as it manages to preserve stable accuracy for the small perturbations.

It is important to note that the envelope of curves per design in Fig.~8 spans about 0.15-0.2 in accuracy scores. This means that the binning design has vast implications on the deterioration rate of the classification scores.

For completeness, we repeat the analysis that was presented in Fig.~8, but for the case of two way category split, e.g., [1-7, 8-10] or [1-5, 6-10]. In this scenario, there are only nine possible splits, and we denote the different configurations by computing the lengths of the top class, i.e., the above two settings correspond to lengths of 3 and 5, respectively. Similar to the findings in Fig.~8, Fig.~9 shows that the best settings are those that have an even split (4 to 6) and that as the splits get more and more uneven, the achievable accuracy decreases even more.

Now, let us examine the results of the CITY survey data where the goal is to explore ways to reduce the accumulation of non-systematic error and move up the curves in Fig.~4.

\begin{figure}[t] 
\centering
\noindent\includegraphics[page=2,width=0.754\columnwidth]{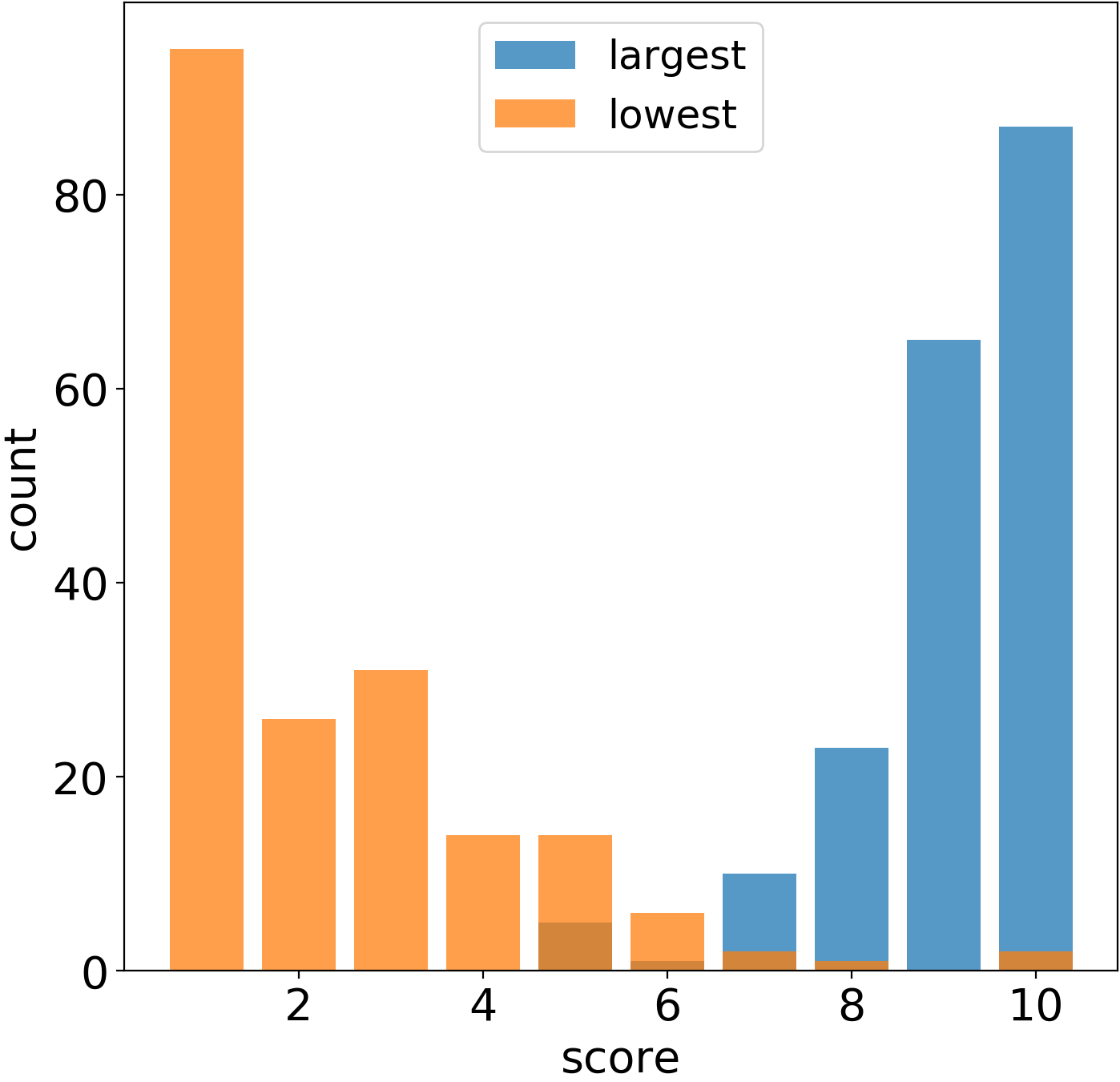}
\caption{The natural highest and lowest scores respondents consider.}
\end{figure}

The CITY survey data allows us to think about the problem from a different, independent perspective.
We surveyed about 200 employees of the retail bank from over 50 different cities spanning over Argentina, China, Hungary, India, Israel, Singapore, the United States, and the UK. As described earlier, people were required to subjectively rate their city as a place of living. 
Figure 10 shows the responses to the question: "Rate your city as a place to live on a scale of 1-10." We can see that most respondents gave their cities high scores of 6 and above. However, no respondents gave a rating below 3. Next, the respondents were asked to assign a category to the numerical score. Figure 10 shows, interestingly, that respondents assign 8-10 to the top category, 4/5-7/8 to the middle one, and 3-4 to the bottom one.

The fact that no respondent chose to give a score below 3 raises the question of whether respondents even considered using the whole spectrum of possible scores. We addressed this question by asking respondents directly what are the highest and lowest scores they would consider giving in a survey like this, and Fig.~11 shows the results. The variability\footnote{computed as the two standard deviations about the mean.} is quite significant: the average highest score stands at $9.1\pm2.3$ while the lowest stands at $2.2\pm3.5$. This variability measures the inter-respondent spread, but still, we argue that, at first approximation, it is a good measure of the intra-respondent variability. Even though this survey is limited in scope and biased toward the retail bank's employees etc., given this approximation and we still would like to mention the resemblance in the variability to the BRAND survey data, which as shown in Fig.~7, has score variability of about $\pm2.5$. 

\begin{figure}[t] 
\centering
\noindent\includegraphics[page=2,width=0.754\columnwidth]{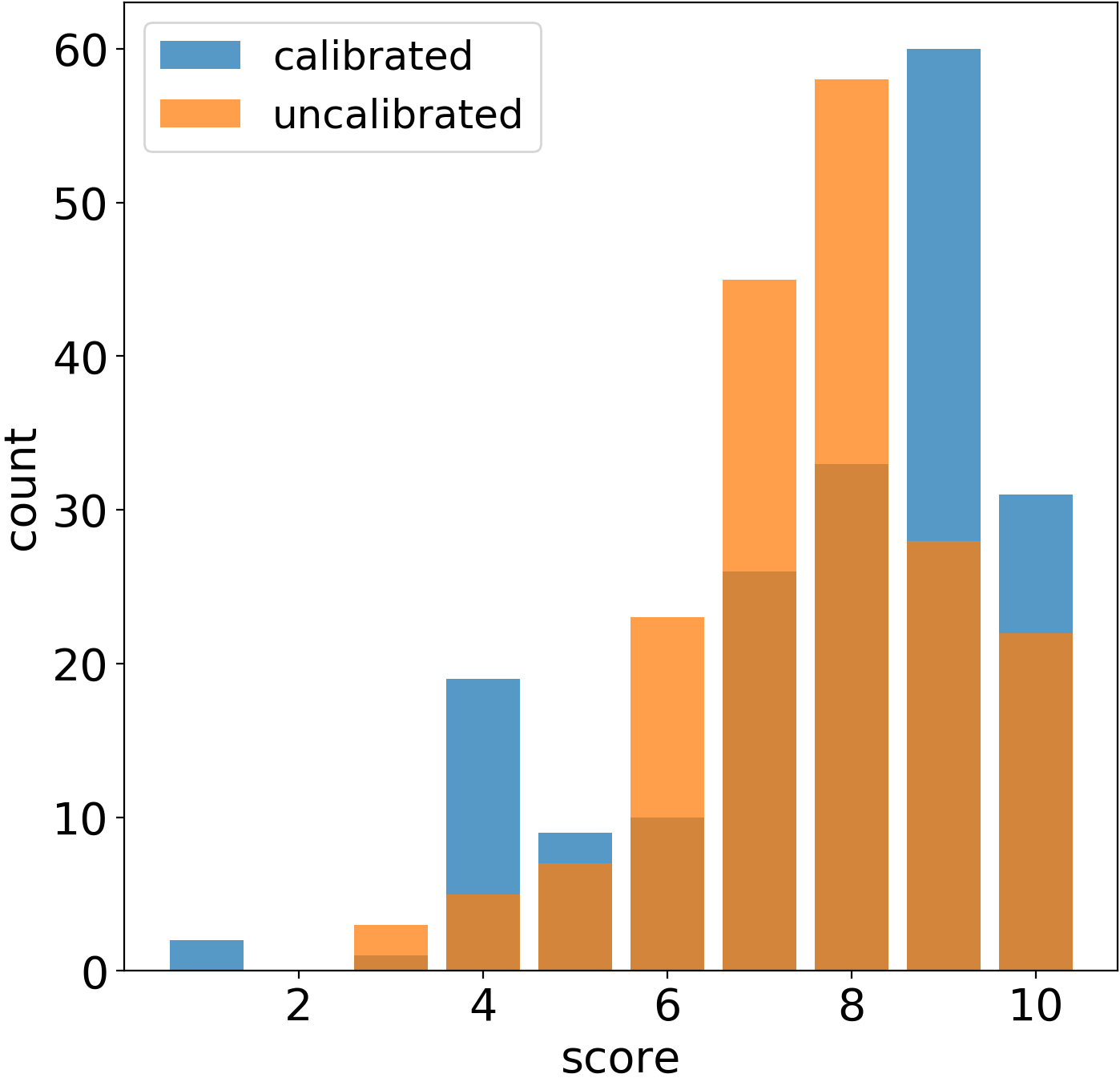}
\caption{Comparing the spread of survey scores in uncalibrated and calibrated surveys.}
\end{figure}

Lastly, Fig.~12 compares the spread of scores in Fig.~10 (denoted by "uncalibrated" survey) to the same question (denoted by "calibrated" survey) except where we attach a short description for each score, as described in the Data and Methods Section. In an example, the choice "9" is replaced by "9) My city is great and I enjoy living in it," and "6" with "6) Its been okay I can't complain," etc. 
Figure 12 shows that in comparison to the uncalibrated survey when we add a short description to the numerical values, we essentially calibrate the survey and as a byproduct, the distribution of responses gets more "uniform" -- survey respondents give responses from a \emph{broader} range of scores. 
To test for uniformity, we apply the Chi-square test to both the uncalibrated and calibrated results that are seen in Fig.~12. 
We find that the calibrated count has p-values 4-orders of magnitude larger, mainly due to the population of the minimum scale. This result indicates that the calibrated survey is closer to uniformity than its uncalibrated counterpart. These findings are robust to both omitting scores less than 3, and when considering the log transformation.

\section{Summary and Conclusions}

In this study, we examine the value that can be extracted from segmented ordinal survey data. We demonstrate that non-systematic error compounds and (almost exponentially) deteriorates classification performance. Further we show that the rate of deterioration can be estimated using simple assumptions, even in real data. 
We observe that, in practice, it is also common to compound this error with another source of error which arises from segmenting the survey responses into unevenly-spaced bins for marketing and business purposes. 
We suggest a simple solution to reduce the non-systematic error. That is, to map each number on the ordinal scale to an appropriate short textual description which calibrates the underlying inconsistent scales of respondents with the survey's rating scale.
Additionally, we suggest avoiding the typical practice of segmenting the results of ordinal surveys. This is made possible by reducing the ordinal scale to only three options: promoters, passives, and detractors. Combining a reduced ordinal survey with short textual description is ideal.
As for plans, we want to examine the deterioration effect using more general metrics as confusion matrices and ROC curves.

\subsubsection*{Acknowledgments}
We thank Jiahao Chen, Robert Tillman, Vamsi Potluru, and Manuela Veloso for important ideas and discussions that helped to ignite this paper. We would also like to thank Christine Cao, Jon Woodward, Carol Liang, and Joe Garcia for insightful comments and important ideas that helped in bringing this manuscript to completion.

\subsubsection*{Disclaimer}
This paper was prepared for information purposes by the Artificial Intelligence Research group of J.~P.~Morgan Chase \& Co.~and its affiliates (“J.~P.~Morgan”), and is not a product of the Research Department of J.~P.~Morgan. J.~P.~Morgan makes no representation and warranty whatsoever and disclaims all liability, for the completeness, accuracy or reliability of the information contained herein, including any data used therein.  All data and other information contained in this material are not warranted as to completeness or accuracy.  This document is not intended as investment research or investment advice, or a recommendation, offer or solicitation for the purchase or sale of any security, financial instrument, financial product or service, or to be used in any way for evaluating the merits of participating in any transaction, and shall not constitute a solicitation under any jurisdiction or to any person, if such solicitation under such jurisdiction or to such person would be unlawful.   

\textcopyright 2020 J.~P.~Morgan Chase \& Co.~All rights reserved.


\bibliographystyle{ACM-Reference-Format}
\bibliography{sample-base}

\end{document}